\documentclass{artificial-life}

\usepackage[style=apa,natbib=true]{biblatex}
\usepackage{amsmath}
\usepackage{amssymb}
\title{What AI Cannot Solve: Rethinking the Foundations of Cryptography}

\auth{
    Yue Zhang \affil{1} \orcid{0000-0002-7786-0231}
}

\corresponding{Yue Zhang}{zyueinfosec@sdu.edu.cn}

\affiliations{
    \item School of Computer Science and Technology,
    Shandong University, Qingdao 266237, China
}

\abs{
Recent advances in artificial intelligence (AI) are challenging
the computational hardness assumptions underlying modern
cryptography. Unlike conventional improvements in computing
power, AI may accelerate the discovery of fundamentally new
mathematical techniques and cryptanalytic algorithms.
However, algorithmically implemented AI remains subject to
the fundamental limits of computation.
In this perspective, we revisit four classical mathematical
results: the undecidability of the halting problem and
Hilbert's tenth problem, the uncomputability of Kolmogorov
complexity, and G\"odel's incompleteness theorems.
We examine their implications for cryptographic design and
identify the gap between general computational impossibility
and the security of efficiently usable cryptographic instances.
We further discuss the mathematical requirements and potential
directions for bridging this gap.
Rather than claiming that undecidability directly guarantees
cryptographic security, we advocate a research agenda towards
post-AI cryptography, exploring whether fundamental limits of
computation can provide new foundations for cryptographic
security beyond conventional hardness assumptions.
}

\begin{document}

% Don't Touch ! %%%%%% 
\coverpage           % 
%\linenumbers        % 
%\doublespacing      %
\onehalfspacing      %
%%%%%%%%%%%%%%%%%%%%%% 

%%% ALL TEXT OF ARTICLE BELOW THIS LINE %%%
%-----------------------------------------%

\section{The AI Challenge to Cryptography}
\label{sec:introduction}

Artificial intelligence (AI) is placing growing pressure on the assumptions underlying modern cryptography. Recent headlines discuss possible cryptographic breakthroughs~\cite{down2026ai}. In late November 2026, this concern was accompanied by the establishment of the Cryptography Incident Review Group, comprising 21 cryptographers who aim to assess cryptanalytic advances and advise on disclosure and resilience~\cite{otto2026cryptographers}. Meanwhile, reporting on alleged attacks against RSA-2048 emphasized that no laboratory had publicly demonstrated such a break~\cite{shivam2026rsa}. These developments highlight the urgency of evaluating AI's implications while distinguishing demonstrated results from speculation.

Published research already provides a concrete basis for this concern. Anthropic reported AI-assisted improvements to attacks against the HAWK signature candidate and a reduced-round variant of AES~\cite{anthropic2026weaknesses}. Although these results did not compromise deployed systems, they demonstrate that AI can contribute to discovering weaknesses in cryptographic designs. The resulting pressure concerns the discovery of algorithms: advances in mathematical reasoning and automated experimentation may reduce attack costs through previously unknown mathematical techniques. Consequently, confidence built through years of cryptanalysis must be continually reassessed as the tools available to researchers and adversaries improve.

We argue that this prospect calls for a closer examination of the foundations on which future cryptographic systems are built. Existing computational security definitions already encompass efficient AI-based adversaries; the concern is whether the hardness assumptions supporting particular schemes remain credible as attack discovery improves. This motivates our central question: \emph{Can fundamental limits of computation provide useful foundations for cryptographic design in the age of AI?} Addressing this question requires understanding what these limits establish and whether they can support mechanisms that combine efficient legitimate computation with quantifiable resistance to attacks.
 
\section{Computational Hardness of Cryptography}
\label{sec:hardness}

Many modern cryptographic schemes, particularly public-key constructions, rely on the assumed hardness of specific computational problems. Their design exploits a computational asymmetry: legitimate users can efficiently perform private-key operations using secret information, whereas adversaries are assumed unable to recover that information from public data within feasible resource bounds. RSA illustrates this asymmetry: generating two large primes and multiplying them is efficient, while recovering those primes from their product is believed to be computationally expensive at suitable parameter sizes~\cite{rivest1978rsa}.

Consider an RSA public key $(N,e)$, where $N=pq$ for distinct secret primes $p$ and $q$. If an adversary can efficiently factor $N$, it can compute
\begin{equation}
    \varphi(N) = (p-1)(q-1).
    \label{eq:rsa-totient}
\end{equation}
Since $\gcd(e,\varphi(N))=1$, the adversary can then obtain a valid private exponent
\begin{equation}
    d \equiv e^{-1} \pmod{\varphi(N)}.
    \label{eq:rsa-private-exponent}
\end{equation}
This exponent enables decryption or signature generation under the corresponding RSA scheme. Consequently, a practical factoring algorithm for deployed RSA moduli directly enables private-key recovery, even when the implementation is otherwise correct~\cite{rivest1978rsa}.

This dependence also appears in post-quantum cryptography. NIST's ML-KEM standard relates its security to the computational difficulty of the Module Learning with Errors (Module-LWE) problem~\cite{nist2024mlkem}. Adopting a different mathematical structure can therefore change the underlying assumptions and attack landscape, but security must still be assessed for the specific problem, instance distribution, parameter sizes, and adversarial capabilities. For AI-assisted cryptanalysis, the consequential question is whether newly discovered algorithms can make these particular attack tasks tractable at the scales used in practice.

\section{Computational Limits of AI}
\label{sec:limits}

An AI whose training, reasoning, and tools are implemented by algorithms remains subject to mathematical limits on computation. These limits concern guarantees across all inputs or within a specified proof system; they do not make every individual instance unsolvable. Four examples illustrate the distinction.

\textbf{The halting problem: predicting whether a program stops.}
Given a program and its input, can an AI always determine whether execution will eventually stop? Suppose such an infallible predictor exists. Construct a program that asks the predictor about its own execution, then runs forever if the answer is ``stops,'' and stops otherwise. Either prediction becomes wrong. This contradiction rules out any algorithm that always terminates with the correct answer for every program~\cite{turing1937computable}. More computing power cannot remove the contradiction.

\textbf{Hilbert's tenth problem: deciding whether integer solutions exist.}
This problem asks whether an arbitrary polynomial equation with integer coefficients has an integer solution, without a preset bound on the variables. Program termination can be encoded into such equations: a corresponding equation has a solution exactly when the encoded program halts~\cite{matiyasevich1971diophantine}. An AI that could decide solvability for every such equation would therefore solve the halting problem. Checking a proposed solution is straightforward, but searching indefinitely does not establish that no solution exists.

\textbf{Kolmogorov complexity: finding the shortest explanation.}
For a fixed universal programming language, the Kolmogorov complexity $K(x)$ measures the length of the shortest program that outputs a string $x$ and halts. This function is uncomputable~\cite{kolmogorov1968information}. Intuitively, finding a short program is different from proving that it is the shortest. A shorter candidate might still be running, and waiting longer does not always reveal whether it will eventually produce the desired output. AI can discover compact descriptions, but cannot always determine the exact minimum description length.

\textbf{G\"odel's incompleteness: limits of a fixed rulebook.}
Suppose an AI proves claims using a consistent set of axioms that can be mechanically listed and is strong enough to express basic arithmetic. The G\"odel--Rosser theorem guarantees statements that can neither be proved nor disproved within that system~\cite{godel1931incompleteness,rosser1936extensions}. Searching faster cannot find a proof that does not exist under those rules. Stronger axioms may settle a particular statement, but any extension satisfying the same conditions will leave further statements undecided.

These results motivate a cryptographic question: can suitable restrictions of these problems yield efficiently usable constructions with a measurable advantage for legitimate users? The impossibility of a general solver does not, by itself, establish the security of the instances used by a cryptosystem.
 
\section{From Computational Limits to Cryptographic Design}
\label{sec:design}

\textbf{Required properties.}
A candidate problem must support efficient generation of a public instance together with a secret witness, efficient legitimate operations using that witness, and hardness on the distribution of instances actually generated. It also needs a connection between witness hardness and message hiding: difficulty in recovering an entire secret does not automatically make every bit of that secret unpredictable.

\textbf{Selecting suitable problems.}
Diophantine equations offer a natural representation of secret solution witnesses. Time-bounded Kolmogorov complexity has an established theoretical connection: specific average-case hardness conditions characterize the existence of one-way functions~\cite{liupass2020kolmogorov}. Halting-inspired candidates would require finite, efficiently verifiable execution witnesses. Proof-based candidates must use statements with available proofs in the chosen system; independent statements cannot supply such proofs.

\textbf{An illustrative encryption construction.}
Consider a public, efficiently computable family of quadratic maps
\begin{equation}
    Q_n:\{0,1\}^{n}\rightarrow\{0,1\}^{\ell(n)}
\end{equation}
over $\mathbb{F}_2$, an established setting for multivariate cryptographic constructions~\cite{berbain2006quad}.
Assume that, for uniformly random $s$, no probabilistic polynomial-time adversary can recover any preimage of $Q_n(s)$ with non-negligible probability. This finite restriction is decidable; its assumed one-wayness does not follow from Diophantine undecidability.

For a single use, Alice and Bob share a uniformly random secret $s\in\{0,1\}^{n}$. To encrypt one bit $m$, Alice independently samples a public random vector $r\in\{0,1\}^{n}$ and computes
\begin{align}
    b &= \left(\sum_{i=1}^{n}s_i r_i\right)\bmod 2,\\
    C &= \bigl(Q_n(s),\,r,\,c\bigr),
    \qquad c=m\oplus b.
\end{align}
Bob recomputes $b$ using $s$ and recovers $m=c\oplus b$.
Under the stated one-wayness assumption, the Goldreich--Levin theorem makes $b$ unpredictable from $(Q_n(s),r)$ with more than negligible advantage over random guessing, providing one-time message hiding.
 
\section{Conclusion}

The rapid advancement of AI calls for a reconsideration of
the mathematical foundations of modern cryptography.
While AI may accelerate the discovery of algorithms that
undermine existing hardness assumptions, it cannot escape
the fundamental limits of algorithmic computation.
These limits offer a largely unexplored perspective on
cryptographic design.
The central challenge is to translate general computational
impossibility into meaningful security guarantees for
efficiently usable cryptographic instances.
We envision post-AI cryptography as a research direction
that investigates whether cryptographic security can be
grounded not only in what adversaries are computationally
unlikely to solve, but also in what algorithmic computation
fundamentally cannot achieve.
\printbibliography

\end{document}